# Tunable metal-insulator transition in double-layer graphene heterostructures

L. A. Ponomarenko[1], A. K. Geim[1,2], A. A. Zhukov[2], R. Jalil[2], S. V. Morozov[1,3], K. S. Novoselov[1], I. V. Grigorieva[1], E. H. Hill[2], V.V. Cheianov[4], V. I. Fal'ko[4], K. Watanabe[5], T. Taniguchi[5], R. V. Gorbachev[2]

[1]School of Physics and Astronomy, University of Manchester, Manchester M13 9PL, UK
[2]Manchester Centre for Mesoscience and Nanotechnology, University of Manchester, Manchester M13 9PL, UK
[3]Institute for Microelectronics Technology, 142432 Chernogolovka, Russia
[4]Physics Department, University of Lancaster, Lancaster LA1 4YB, UK
[5]National Institute for Materials Science, 1-1 Namiki, Tsukuba, 305-0044 Japan

*We report a double-layer electronic system made of two closely-spaced but electrically isolated graphene monolayers sandwiched in boron nitride. For large carrier densities in one of the layers, the adjacent layer no longer exhibits a minimum metallic conductivity at the neutrality point, and its resistivity diverges at low temperatures. This divergence can be suppressed by magnetic field or by reducing the carrier density in the adjacent layer. We believe that the observed localization is intrinsic for neutral graphene with generic disorder if metallic electron-hole puddles are screened out.*

Disordered conductors with resistivity $\rho$ above the fundamental value of the resistance quantum $h/e^2$ are expected to exhibit an insulating behaviour at low temperatures ($T$), that is, their $\rho$ should diverge as $T \rightarrow 0$ [1-3], in contrast to the metallic behavior with $\rho$ decreasing and saturating at low $T$. In essence, the above value of $\rho$ indicates that the electron mean free path $l$ is shorter than the Fermi wavelength $\lambda_F$ so that quantum interference becomes a dominant feature in electrons' diffusion, leading to a strong (Anderson) localization. The phenomenon has been observed in a multitude of materials, including damaged graphene and its disordered chemical derivatives, [4-10] and its scope extends beyond electronic systems, into optical and acoustic phenomena too. Surprisingly, no sign of Anderson localization has been observed in generic graphene that remains metallic at liquid-helium $T$ [3-5]. This is despite the fact that, near the neutrality point (NP), it has $\rho \approx h/e^2$ per carrier type and, therefore, is close to the onset of Anderson localization. Generic graphene exhibits only a relatively weak $T$-dependence that can be explained by phonons and thermally excited carriers [4,11]. Earlier theoretical studies have suggested that Dirac electrons can evade localization for certain types of disorder [3,12-15], with the extreme example being graphene subjected to a smooth Coulomb potential [16,17]. However, for generic disorder that involves scattering between the two graphene valleys, the localization is expected to be unavoidable [3,18,19]. It has remained a puzzle why this does not happen in experiment.

In this Letter, we report Anderson localization in high-mobility graphene devices (carrier mobility $\mu \sim 100,000$ cm$^2$/Vs), which exhibit a metal-insulator transition (MIT) with increasing rather than decreasing their quality and homogeneity. The transition can be controlled externally, by a second graphene layer placed at a distance of several nm and isolated electrically. In the following, the two layers in the double layer graphene (DLG) heterostructure are referred to as studied and control. At low doping $n_C$ in the control layer, the studied layer exhibits the standard behavior with a minimum metallic conductivity of $\sim 4e^2/h$. However, for $n_C > 10^{11}$cm$^{-2}$, resistivity $\rho$ of the studied layer diverges near the NP at $T<70$K. This divergence can be suppressed by a small perpendicular field $B <0.1$T, which indicates that this is an interference effect rather than a gap opening. We attribute the MIT to the recovery of an intrinsic behavior such that graphene exhibits Anderson localization if its $\rho$ reaches values of $\approx h/e^2$ per carrier type. We believe that, normally, this intrinsic MIT is obscured by charge inhomogeneity in the form of electron-hole (e-h) puddles [4,20-24]. Within each puddle, graphene is sufficiently away from the NP and remains metallic. Then, resistivity of the percolating e-h system with leaking p-n boundaries [16,17] assumes a value of $\sim h/e^2$ with little $T$ dependence (note that conceptually this value has little in common with the similar value required for Anderson localization) [23,24]. The control layer can screen out the fluctuating background potential and suppress e-h puddles, revealing the intrinsic insulating properties at the NP. This reconciles the metallic behavior normally observed in graphene at low $T$ with the localization expected for such large $\rho$ and supports the idea that the minimum conductivity that tends to assume values close to $4e^2/h$ is due to e-h puddles [23,24].

The studied devices were fabricated by sandwiching two graphene monolayers with thin hexagonal-BN crystals. In a multistep procedure described in the supplement, a graphene monolayer was transferred onto a 20-30 nm thick BN crystal that was first prepared on top of an oxidized Si wafer. Then, the graphene was covered by another BN crystal (spacer), which followed by transfer of the second graphene layer. Both layers were shaped into multiterminal devices aligned above each other and having separate electrical contacts (Fig. 1a). Individual steps were similar to those described in [25,26] but the whole fabrication process involved 3 dry transfers and alignments, 4 nonconsecutive rounds of electron-beam lithography, 3 rounds of plasma etching and two separate metal depositions. The resulting DLG heterostructures are schematically shown in Fig. 1a (also, see the supplement). We made several such devices with channel widths of 1 to 2 μm. They exhibited μ of $30 \div 120 \times 10^3$ cm$^2$/Vs and little chemical doping. The bottom layer encapsulated in BN always had higher μ and changed little after exposure to air [26] ] whereas quality of the top layer gradually decayed. For this particular study, we employed three multiterminal devices with sufficiently thick BN spacers to avoid any detectable tunnel current between graphene layers (<0.1 nA). The spacers had thickness $d$ of ≈4, 12 and 16 nm. All the devices exhibited a similar MIT behavior, although the insulating state was much more pronounced for devices with smaller $d$ and higher μ as described below.

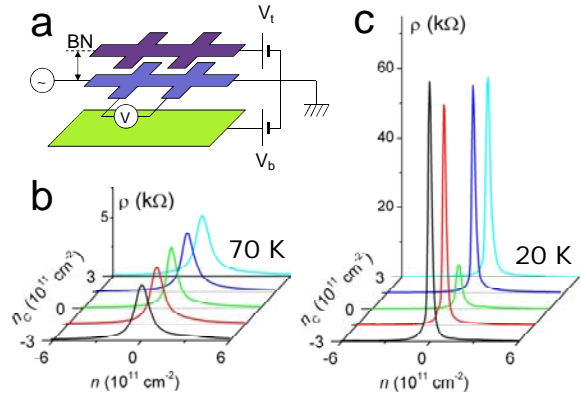

FIG. 1 (a) – Schematic view of our heterostructure devices and measurement geometry. (b,c) – ρ as a function of $n$ in the studied graphene layer for different doping $n_C$ of the control layer at two temperatures. The device has a hexagonal-BN spacer with $d$ =4 nm.

With reference to Fig. 1a, we employed the following scheme of measurements. Voltage $V_t$ was applied between the two graphene layers, and this electrically doped both of them with carriers of the opposite sign. The bottom layer could also be gated by voltage $V_b$ applied to the Si wafer. Because of the low density of states, graphene can provide only a partial screening and, therefore, $V_b$ induced carriers in the top layer as well. This influence was weaker than on the bottom layer and depended on $n$ in the latter. By measuring Hall resistivity $\rho_{xy}$ we could determine $n$ in each of the layers (supplement). We usually fixed $V_t$ to define a nearly constant $n$ in the top layer and swept $V_b$ to vary $n$ in the bottom layer. Normally, we studied the higher-μ bottom layer and used the top layer as control. In this configuration, the insulating state reached higher ρ. If the studied and control layers were swapped, the behavior remained qualitatively the same (see the supplement) but lower μ resulted in lower ρ of the insulating state.

Our main result is illustrated by Fig. 1 that plots two sets of standard curves ρ($n$) for the studied layer at different $n_C$. At 70K, the control layer causes little effect on the studied layer, and all the curves in Fig. 1b look no different from those observed in the standard devices [4] or for graphene on BN (GBN) [25]. However, at low $T$ and for high doping of the control layer ($n_C >10^{11}$cm$^{-2}$), graphene exhibits a radically different behavior (Fig. 1c). In this regime, ρ at the NP acquires a strong $T$ dependence and easily overshoots the threshold value of $h/e^2$. To elucidate this observation, Fig. 2 shows further examples of ρ($n,T$) for high and low doping of the control layer. In the case of large $n_C$ (Fig. 2a), ρ exhibits an insulating $T$ dependence. The behavior is in contrast with the weaker $T$ dependence for zero $n_C$ (the latter can be explained by thermally excited carriers) (Fig. 2b). Outside a relatively narrow interval, $|n| \le 10^{10}$cm$^{-2}$ and above 70K, graphene's behavior was practically independent of $n_C$.

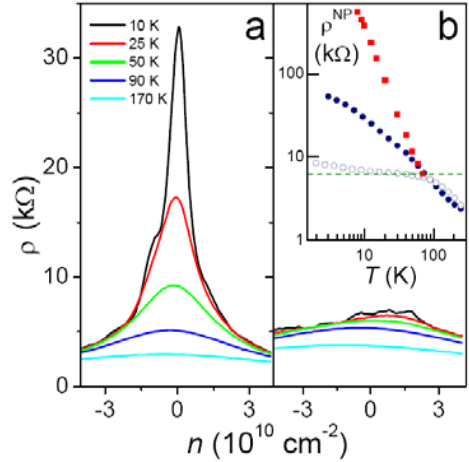

FIG. 2. Resistivity of the studied layer at different $T$ for high and low doping of the control layer. (a,b) correspond to $n_C \approx 3 \times 10^{11} cm^{-2}$ and zero $n_C$, respectively. Here, we have chosen to plot data for $d \approx 12$ nm. For our thinnest spacer ($\approx 4$ nm), $\rho^{NP}$ becomes very large at low $T$ (inset) and continuous curves $\rho(n)$ are difficult to measure because of crosstalk nonlinearities (supplement). The inset shows the $T$ dependence of $\rho^{NP}$ for the device in the main figure at both $n_C$ (open and solid circles) and for the 4 nm device at $n_C \approx 5 \times 10^{11} cm^{-2}$ (squares). The dashed line indicates the standard threshold value for Anderson localization with 4 carrier flavors, $\rho = h/4e^2$.

The $T$ dependence of the maximum resistivity at the NP, $\rho^{NP}$, is shown in more detail in the inset of Fig. 2b for zero- and heavily- doped control layers. The insulating state is more pronounced for $d =4$ nm but remains clear also for the 12 nm device (note the logarithmic scale). For $d =4$ nm and below 4K, $\rho^{NP}$ could reach into the MΩ range (supplement), an increase by 2-3 orders of magnitude with respect to the standard behavior. The high-$\rho$ regime is found to be difficult to probe due to a strong nonlinearity caused by a crosstalk between the measurement current and $V_t$, the effect specific to DLG devices (supplement). To assure the linear response in this regime, we had to measure the I-V curves at every gate voltage (supplement) and, to avoid these difficulties, we limited our studies mostly to $T$ >4K and $\rho^{NP}$ <100 kΩ.

The influence of the adjacent layer immediately invites one to consider interlayer Coulomb interactions. Indeed, the relevant energy scale is $e^2/\varepsilon d$ ~50 meV, that is, the interactions may be a significant factor ($\varepsilon \approx 5$ is BN's dielectric constant). For example, one can imagine that the interactions open an excitonic-like gap at the Dirac point. We have ruled out this possibility by magnetic field measurements. In the gapped case, $B$ is expected to enhance the confinement and, hence, the binding energy. In contrast, our devices exhibit a pronounced negative magnetoresistance in non-quantizing $B$ (Fig. 3). The insulating behavior is suppressed in characteristic $B^*$ $\approx$10mT (Fig. 3), well below the onset of Landau quantization. Fig. 3 also shows that the MIT is again confined to a region $|n| \leq 1 \times 10^{10} cm^{-2}$.

Another revealing observation is that the insulating state in our experiments has always developed at $\rho > h/4e^2$ (Figs. 1-3). This is seen most clearly in the inset of Fig. 2 where the curves depart from each other above the dashed line marking $h/4e^2$. In the insulating state, $\rho^{NP}$ is found to follow a power-law dependence $1/T^\nu$ where $\nu$ varied from sample to sample reaching a value close to 2 in the device with $d \approx$4nm. The characteristic $T$ at which the insulating state started to develop can be attributed to the fact that above 70K the concentration of excited carriers at the NP exceeded $\approx 10^{10} cm^{-2}$, beyond which no MIT could be observed even at low $T$.

The suppression of the MIT by non-quantizing $B$ is a clear indication that localization plays an important role, such that $B$ breaks down the time-reversal symmetry and destroys the interference pattern due to self-intersecting trajectories [1-3]. The strong localization scenario is also consistent with the onset of the insulating state at $\rho^{NP}$ $\approx h/4e^2$, which corresponds to the resistivity quantum per carrier type. However, localization in graphene cannot possibly be explained without intervalley scattering [3,18,19]. A tempting line of arguing would be to invoke charge fluctuations in the control layer to explain its influence on the studied layer. However, this contradicts to the fact that µ can notably increase at high $n_C$, that is, graphene exhibits higher quality rather than extra scattering if the control layer is strongly doped (supplement). Moreover, the Coulomb interaction between the layers is generally expected to become less efficient with decreasing $T$ and increasing $n_C$ [27], which is exactly opposite to what we observe. Finally, an interlayer scattering mechanism can be ruled out by the fact that any interaction potential created by carriers in the control layer and acting on the studied one varies at distances of $\sim d \gg a$ ($a$ is the lattice constant) whereas the fast components needed for intervalley scattering depend exponentially on $a/d$ [28].

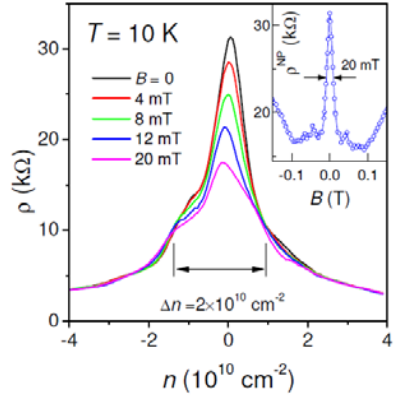

FIG. 3. Resistivity of the studied layer in the insulating regime at various $B$. $d = 12$ nm; $n_C \approx 3 \times 10^{11}$ cm$^{-2}$. At low $T$, $\rho(n)$ exhibits pronounced mesoscopic fluctuations (e.g., the left shoulder in this figure), which develop further with decreasing $T$. Inset – detailed $B$ dependence of $\rho^{NP}$ for the case shown in the main figure.

To explain the MIT, we assume a small amount of intervalley scatterers already present in our devices. They could be either some of the defects that limit µ (e.g., strong adsorbates) [29] or, alternatively, the intervalley scattering can arise due to the atomic-scale potential created by BN. In both cases, this can break down the symmetry between the carbon sublattices and act as a source of intervalley scattering. . Because the insulating state is observed only for $|n| \leq 10^{10}$ cm$^{-2}$ and the process responsible for Anderson localization requires the scattering length of about $\lambda_F \sim n^{-1/2}$, we can estimate the intervalley scattering length $l_{iv}$ as $\sim 0.1$ µm.

Furthermore, $B^* \sim 10$ mT yields a spatial scale $(\phi_0/B^*)^{1/2} \approx 0.5$ µm, which corresponds to a flux quantum $\phi_0 = h/e$ enclosed by diffusive trajectories. This scale is significantly larger than the mean free path $l \leq 0.1$ µm estimated for the relevant interval of $n \leq 10^{10}$ cm$^{-2}$ and, therefore, this justifies the use of diffusive transport notions. By fitting the magnetoresistance curves such as in Fig. 3 by the weak localization formulas [19] (though mentioning that those are applicable to small rather than large changes in ρ) yields two other spatial scales. One corresponds to the onset of magnetoresistance (~1mT) and yields the phase-breaking length of a few µm at liquid-helium $T$, which is typical for graphene [30]. The other scale (≈0.1µm) is given by $B \approx 0.1$T where the magnetoresistance saturates, before changing its sign from negative to positive. The latter scale can be due to the onset of intervalley scattering [18,19,30], which agrees well with the value $l_{iv}$ determined from the above analysis of the MIT.

The proposed scenario for the MIT could be considered routine for any high-ρ metallic systems at low $T$. So far, graphene has been the only known exception, unless intentionally damaged [6-10]. Therefore, the question should be turned around and asked why there is no MIT in the standard graphene devices or DLG at low $n_C$, and why the MIT becomes more pronounced in our ultra-high-quality graphene. The latter seemingly contradicts the very notion of Anderson localization. The puzzle allows a straightforward resolution if we attribute this behavior to the presence of e-h puddles [4,23,24].

In graphene on $SiO_2$, the puddles contain carries in typical $n \sim 10^{11}$ cm$^{-2}$ [20]. In GBN, puddles are larger and shallower [21,22] but, within each puddle, $n$ is still large enough (>$10^{10}$ cm$^{-2}$) to move the system away from the MIT. Resistivity of such an inhomogeneous system is then determined by inter-puddle ballistic transport with ρ $\sim h/4e^2$ [23,24]. The recovery of the MIT can be expected if $n$ within e-h puddles decreases below the localization threshold ($\approx 10^{10}$cm$^{-2}$ in our case). Accordingly, we attribute the influence of the control layer to the fact that at high $n_C$ it screens out the background potential, making puddles shallower. Experimentally, this is indeed the case, as seen from Hall measurements where the transition region in $\rho_{xy}$ (between e- and h- doping) notably narrows at high $n_C$ (supplement).

In summary, an insulating state is found near the NP in high-µ graphene with low charge inhomogeneity. The observed behavior can be explained by the suppression of e-h puddles that disallow Anderson localization in standard-quality graphene. The MIT that occurs with decreasing disorder is a unique and counterintuitive phenomenon such that further work is required to better understand the underlying physics, especially, the mechanism of intervalley scattering and a possible role of the atomic washboard created by BN.

REFERENCES
[1] B. Kramer, A. MacKinnon, *Rep. Prog. Phys.* **56**, 1469 (1993)
[2] M. Imada, A. Fujimori, Y. Tokura, *Rev. Mod. Phys.* **70**, 1039 (1998)
[3] F. Evers, A. D. Mirlin, *Rev. Mod. Phys.* **80**, 1355 (2008)
[4] A. K. Geim, K. S. Novoselov, *Nature Mat.* **6,** 183 (2007)
[5] A. H. Castro Neto *et al*, *Rev. Mod. Phys.* **81,** 109 (2009)
[6] C. Gomez-Navarro, T. R. Weitz, A. M. Bittner, M. Scolari, A. Mews, M. Burghard, K. Kern, *Nano Lett.* **7**, 3499 (2007)
[7] S. Y. Zhou, D. A. Siegel, A. V. Fedorov, A. Lanzara, *Phys. Rev. Lett.* **101,** 086402 (2008).
[8] A. Bostwick, J. L. McChesney, K. V. Emtsev, T. Seyller, K. Horn, S. D. Kevan, E. Rotenberg, *Phys. Rev. Lett.* **103**, 056404 (2009)
[9] D. C. Elias *et al*, *Science* **323**, 610 (2009)
[10] J. H. Chen, W. G. Cullen, C. Jang, M. S. Fuhrer, E. D. Williams, *Phys. Rev. Lett.* **102**, 236805 (2009)
[11] S. Adam, S. Das Sarma, *Phys. Rev. B* **77**, 115436 (2008).
[12] A.A. Nersesyan, A. Tsvelik, F. Wegner, *Phys. Rev. Lett.* **72**, 2628 (1994)
[13] A. Ludwig et al, *Phys. Rev. B* **50**, 7526 (1994)
[14] Y. Hatsugai, X.-G. Wen, M. Kohmoto, *Phys. Rev. B* **56**, 1061 (1997)
[15] S. Ryu and Y. Hatsugai, *Phys. Rev. B* **65**, 033301 (2001)
[16] M. I. Katsnelson, K. S. Novoselov, A. K. Geim, *Nature Phys.* **2,** 620 (2006)
[17] V.V. Cheianov and V.I. Fal'ko, *Phys. Rev. B* **74**, 041403 (2006)
[18] I. L. Aleiner, K.B. Efetov, *Phys. Rev. Lett.* **97**, 236801 (2006).
[19] E. McCann, K. Kechedzhi, V. I. Fal'ko, H. Suzuura, T. Ando, B. L. Altshuler, *Phys. Rev. Lett.* **97**, 146805 (2006)
[20] J. Martin, N. Akerman, G. Ulbricht, T. Lohmann, J. H. Smet, K. von Klitzing, A.Yacoby, *Nature Phys.* **4**, 144 (2008)
[21] J. Xue *et al*, *Nature Mat.* **10,** 282 (2011)
[22] R. Decker *et al*, *Nano Lett.* **11**, 2291 (2011)
[23] S. Adam, E. H. Hwang, V. M. Galitski, S. Das Sarma, *Proc. Ntl. Acad. USA* **104**, 18392 (2007)
[24] V. V. Cheianov, V. I. Fal'ko, B. L. Altshuler, I.L. Aleiner, *Phys. Rev. Lett.* **99,** 176801 (2007)
[25] C. R. Dean *et al*, *Nature Nano.* **5**, 722 (2010)
[26] A. S. Mayorov *et al*, *Nano Lett.* **11**, 2396 (2011)
[27] W. K. Tse, B. Y. K.Hu, S. Das Sarma, *Phys. Rev. B* **76**, 081401 (2007)
[28] L. A. Ponomarenko *et al*, *Phys. Rev. Lett.* **102**, 206603 (2009)
[29] Z. H. Ni *et al, Nano Lett.* **10**, 3868 (2010)
[30] F. V. Tikhonenko, A. A. Kozikov, A. K. Savchenko, R. V. Gorbachev, *Phys. Rev. Lett.* **103**, 226801 (2009)

**Supplementary Material: Tunable metal-insulator transition in double-layer graphene heterostructures**
L. A. Ponomarenko *et al.*

#1 *Sample fabrication*

The studied devices (see Fig. S1) were fabricated by using the following multistep procedure. First, relatively thick (20-30 nm) crystals of hexagonal BN (hBN) were cleaved on top of an oxidized Si wafer ($SiO_2$ had thickness of either ≈90 or ≈300 nm), and graphene were prepared by cleavage on another substrate covered with PMMA. Then, the chosen graphene crystal was transferred on top of the chosen hBN crystal. To this end, we used alignment procedures similar to those described in ref. [S1,S2]. Electron-beam lithography and oxygen plasma etching were employed to define a 10-terminal Hall bar (Fig. S1). The second, thinner hBN crystal (thickness $d$) was carefully aligned to encapsulate the Hall bar but leave the contact regions open for a later deposition of metal contacts. After this, the second graphene crystal was prepared and transferred on top of the encapsulated graphene device. This second layer was also patterned into another Hall bar device that was carefully aligned with the bottom graphene structure (with a typical accuracy of ~10 nm). Finally, Au/Ti contacts we fabricated by using e-beam lithography and evaporation. After each transfer step, the devices were annealed at 300°C in an argon-hydrogen mixture to remove polymer residues and other contaminants. Figure S1 shows an optical image of a completed DLG device with $d \approx 12$nm.

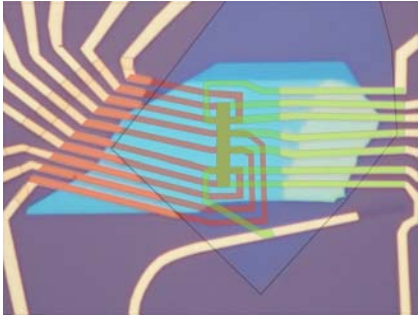

Figure S1. Double-layer graphene heterostructures. We use false colors to identify different layers in the heterostructure. The bottom graphene mesa is shown in false orange and lies on top of an opaque BN crystal (cyan to white is its natural color). A 12 nm hBN crystal resides on top of the graphene layer (transparent blue is its natural color but we added a thin black contour to identify the BN crystal's position). The top graphene layer is shown in false green. It is etched into the multiterminal Hall bar aligned with the bottom mesa. The width of the Hall bar is ≈2μm.

The exact thickness of the BN spacer was found retrospectively by using atomic force microscopy and capacitance measurements [S3]. Both techniques yielded the same value of $d$, which indicates the absence of any residue between the graphene layers (an extra monolayer of contamination would be detectable by this approach). Let us also stress the excellent insulating quality of hexagonal BN [S4], which allows a dielectric layer with $d \leq 4$nm without any noticeable leakage current for gate voltages up to several V applied between the graphene layers [S2]. Such a separation between two-dimensional electronic systems is difficult if not impossible to achieve for GaAlAs heterostructures [S5].

#2 *Relation between gate voltages and induced carrier density*

In field-effect devices with a thick dielectric layer (e.g., using graphene on $SiO_2$), one can generally assume a linear relation $n \propto V_g$ between the induced carrier concentration $n$ and gate voltage $V_g$. However, in our DLG heterostructures with an ultra-thin BN spacer the dependence becomes nonlinear due to quantum capacitance (QC) [S3]. Its contribution becomes particularly important in high-μ devices, in which one can approach close to the NP so that the density of states in graphene tends to zero [S3]. In our case of a double layer system, one needs to take into account that QC contributions come from both graphene layers, and this makes the relation between carrier densities in the control and studied layers ($n_C$ and $n_S$, respectively) and applied voltages $V_t$ and $V_b$ particularly complicated and deserving a separate study.

To simplify our analysis in this work, we employed a constant capacitance approximation. To this end, we fixed $V_t$ that induced charge carriers in the control layer, and determined $n_C$ by using Hall measurements. Then, we swept $V_b$ to change $n_S$ in the studied layer (in the main text and below, we use notation $n$ instead of $n_S$, unless this causes confusion). To convert $V_b$ into $n_S$, we again measured the Hall effect away from the NP, typically at $|n_S| \approx 1$ to $5 \times 10^{11} cm^{-2}$. The inferred coefficient was used to translate $V_b$ into $n_S$. The coefficient changed with varying $n_C$ in the control layer. The latter was taken into account when the experimental curves were re-plotted in terms of $n$ in the figures presented in this work.

The "linear approximation" assumes that QCs do not change significantly within the studied interval of $n$ for a given $n_C$. As explained above, this is a simplification for a complex dependence of $n_C$ and $n_S$ as a function of $V_t$ and $V_b$ which is unique for every device [S3]. The approximation leads to deviations from the actual values of carrier concentrations in the two layers. We have found that deviations are relatively minor (typically, <20% for our range of studied $n$) and become significant only in the proximity of the NP where graphene's QC is minimal. In this regime, charge inhomogeneity is also significant (e-h puddles) which leads to a leveling-off of the decrease in QC. Furthermore, near the NP, we cannot determine $n$ from Hall measurements because of charge inhomogeneity [S3], which makes it difficult if not impossible to improve further on the used linear approximation.

Despite some drawbacks, our approach is more meaningful than just quoting applied voltages that strongly vary for various devices, measurement configurations, etc. The linear approximation has no impact on any of the reported results, and all the curves remain qualitatively the same. However, if detailed analysis is needed, one has to keep in mind that the carrier densities $n$ plotted in the main text are in fact scaled gate voltages, and there is some nonlinearity along the x-axis, which increases near the NP. When a better approximation is required (for example, to find μ; see below), we measured the Hall effect for every value of $V_b$ to find $n$ away from the NP.

#### #3 Crosstalk between measurement current and effective gate voltage

In DLG devices with nm-thin spacers, special care should be taken to avoid the influence of measurement current $I$ on the induced carrier densities. Even for seemingly small currents $I$ ~1nA, a voltage drop along the graphene device can reach into the mV range in the low-$T$ insulating regime with high ρ (in the MOhm range). With reference to Fig. 1a, this voltage drop translates into an additional gate voltage $\Delta V_t$ that varies along the device. Unlike for the standard graphene devices with a thick dielectric, in our devices the effect of such an extra $V_t$ is not negligible because of small $d$. For example, if $d \approx 4$ nm, 1mV of $V_t$ translates into ~$5 \times 10^9$ cm$^{-2}$, which is enough to shift the system away from the insulating state. For example, this can lead to an artifact of the resistance peak split into two. The crosstalk makes measurements in the insulating regime particularly difficult (Fig. S2).

To avoid artifacts in determining $\rho^{NP}(T)$ (insert in Fig. 2) we measured $I$-$V$ curves at the NP at each $T$ (see Fig. S2). Then, $\rho^{NP}$ was defined from the linear part at $I \Rightarrow 0$. Because the range of the linear response shrinks with increasing $\rho^{NP}$, the crosstalk was probably responsible for some rounding of the curves at low $T$, which is seen in the inset of Fig. 2b. This is why we avoided the regime of low $T$ and very high $\rho^{NP}$.

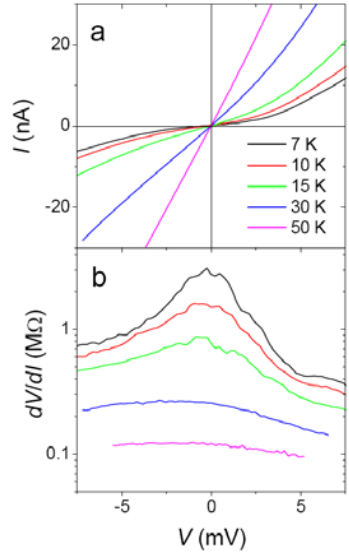

FIGURE S2. I-V characteristics in the insulating state for a device with $d$ =4 nm. The nonlinearity in this state is mostly due to the crosstalk between the driving current and gate voltage. This was confirmed by studying changes in ρ($n$) with increasing $I$ in different parts of the multiterminal devices, which caused different shifts of the NP as a local value of gate voltage varied by a few mV. (b) – Differential resistance for the curves in (a).

#### #4 Suppression of electron-hole puddles by doping of a nearby graphene layer

The suppression of e-h puddles in the studied graphene layer, when the control layer is set in a highly doped state, is an important notion that we used to explain the MIT transition. Although the idea is rather intuitive, we could confirm the suppression directly in an experiment. To this end, we monitored charge inhomogeneity in our DLG devices. The extent of the region with e-h puddles is often characterized by the width of the $\rho(V_g)$ peak [S6]. Our curves indeed become noticeably narrower at high doping of the control layer. However, to elucidate the broadening in more detail, it is useful to employ Hall measurements (Fig. S3). If graphene has only one type of charge carriers, its Hall resistivity follows the dependence $\rho_{xy} = 1/neB$ for both electrons and holes. The transition regime in which electrons and holes coexist (that is, e-h puddles coexist) corresponds to the region around the NP where $\rho_{xy}$ as a function of electric doping changes its sign and reaches a maximum value, before following the $1/n$ dependence (Fig. S3). The width of this region is a good measure of charge inhomogeneity. The maximum value of $\rho_{xy}$ is another way to judge the extent of e-h puddles' region. For graphene on hBN, the region of the coexistence of electrons and holes usually extends to several$\times 10^{10}$ cm$^{-2}$ [S7] but this value corresponds to the disappearance of the deepest puddles. An onset of the MIT should probably be expected at lower $n$ because of the required percolation.

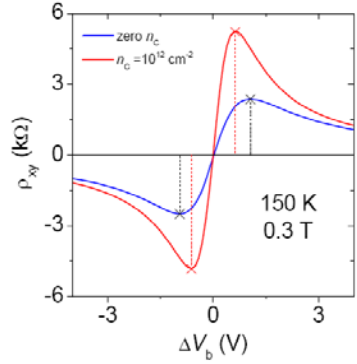

FIGURE S3. Typical changes in Hall resistivity of the studied layer with doping of the control layer. If the control layer is in its undoped state, the studied layer exhibits the behavior shown by the blue curve. Other examples of such curves and their analysis can be found in ref. [S7]. In the strongly doped regime ($n_C \approx 10^{12}$ cm$^{-2}$), the Hall curves become markedly sharper and the transition narrower (typically, by a factor of 2), which translates into twice shallower puddles. The data are for $d = 12$ nm.

#5 *Reciprocity between top and bottom layers*

Although we have normally studied the bottom layer that was encapsulated in BN and showed high $\mu$, the MIT could also be realized, if we swapped the studied and control layers. Fig. S4 shows an example of the MIT in the lower-$\mu$ top layer as a function of doping of the bottom layer. The transition is much less pronounced but, clearly, $\rho^{NP}$ becomes larger with increasing $|n_C|$. This behavior is in agreement with our model for the MIT, that is, lower $\mu$ translates into deeper puddles that are harder to suppress by external screening.

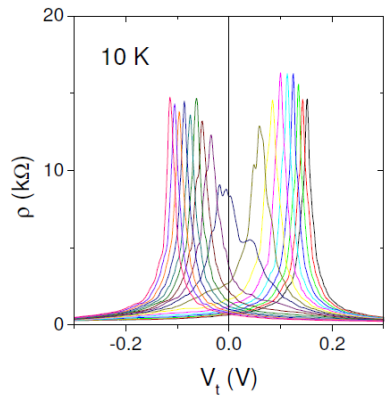

FIGURE S4. Resistivity of the low-$\mu$ top layer as a function of gate voltage $V_t$ for different $n_C$ in the bottom layer. $n_C$ was varied by changing $V_b$ in steps of 2V between -15 and +17 V (curves of different color). The device has the 4 nm spacer.

#6 *Influence of external screening on charge carrier mobility*

The scattering mechanisms that limit charge carrier mobility in graphene remain debated and probably vary for different devices and substrates. Due to the possibility to partially screen out the Coulomb scattering potential in our DLG heterostructures, we can prove that there is more than one type of scatterers, at least, in our devices. For high-$\mu$ graphene (usually, the bottom layer), we find that µ significantly increases if a high density is induced in the top layer (upper panel in Fig. S5). This yields a significant role of Coulomb scattering in such graphene on hBN. On the other hand, for lower-µ graphene (usually, the top layer), we find little effect of $n_C$ on µ, which suggests a non-Coulomb scattering mechanism.

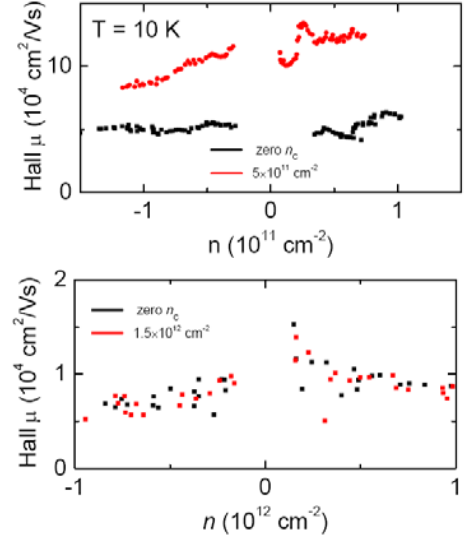

FIGURE S5. Changes in carrier mobility in the studied layer with doping of the control layer. High-µ layer is clearly sensitive to $n_C$, and its mobility increases from ~50,000 to 100,000 cm$^2$/Vs (top panel). No such changes were observed for low-µ graphene (µ~10,000 cm$^2$/Vs; low panel). In this case, we have chosen to present the data for the lowest µ observed in our DLG devices, which is similar to values for graphene on SiO$_2$. After several exposures to air and consecutive annealing, quality of graphene in the top layer gradually decayed from original µ ≥30,000 cm$^2$/Vs. The bottom layer was much more stable. Also, note that due to changes in quantum capacitance, the usual way of determining µ from the electric field effect (as $\sigma = ne\mu$) become unreliable and leads to significant errors in DLG devices (because $n$ is no longer a linear function of gate voltage). Therefore, for better accuracy, we have determined $n$ from Hall measurements in small $B$, which yields the Hall mobility instead of the field-effect one.

The fact that the MIT in our devices is accompanied by a pronounced increase in µ (therefore, in the mean free path $l$) clearly distinguishes our observation from the conventional MIT. To the best of our knowledge, Anderson localization has never been reported with decreasing disorder. Such behavior is counterintuitive but consistent with the proposed model for the MIT.

*Supplementary references*

[S1] C. R. Dean *et al*, *Nature Nano.* **5**, 722 (2010).
[S2] A. S. Mayorov *et al*, *Nano Lett.* asap online (2011).
[S3] L. A. Ponomarenko *et al*, *Phys. Rev. Lett.* **105**, 136801 (2010).
[S4] T. Taniguchi, K. Watanabe. *J. Crystal Growth* **303**, 525 (2007).
[S5] A. F. Croxall, K. Das Gupta, C. A. Nicoll, M. Thangaraj, H. E. Beere, I. Farrer, D. A. Ritchie, M. Pepper, *Phys. Rev. Lett.* **101**, 246801 (2008); J. A. Seamons, D. R. Tibbetts, J. L. Reno, M. P. Lilly, *Appl. Phys. Lett.* **90**, 052103 (2007).
[S6] K. S. Novoselov *et al*, *Nature* **438**, 197 (2005).
[S7] D. A. Abanin, R. V. Gorbachev, K. S. Novoselov, A. K. Geim, L. S. Levitov. arXiv:1103.4742.